\documentclass{aa}

\usepackage{graphicx}
\usepackage{psfrag}
\usepackage{pifont}

\newcommand{\od}[1]{\mbox{${\cal O}(#1)$}}
\def\Div{\mathop{\hbox{div}}\nolimits}
\newcommand{\ntau}{\Pisymbol{psy}{116}}
\newcommand{\infapp}{\raisebox{-.7ex}{$\stackrel{<}{\sim}$}}

\begin{document}

\title{On Jupiter's inertial mode oscillations}

\author{Boris Dintrans\inst{1,2}\fnmsep\thanks{Supported by the European
Commission under Marie-Curie grant no.\ HPMF-CT-1999-00411} \and Rachid
Ouyed\inst{1}}

\institute{Nordic Institute for Theoretical Physics, Blegdamsvej 17,
DK-2100 Copenhagen, Denmark \and Laboratoire d'Astrophysique de Toulouse,
Observatoire Midi-Pyr\'en\'ees, 14 avenue E. Belin, 31400 Toulouse,
France}

\offprints{dintrans@nordita.dk}

\date{Received 21 June 2001/Accepted 8 July 2001}

\abstract{ 
Properties of inertial modes of Jupiter are investigated for an $n=1$
polytropic description of the planet interior. We use the anelastic
approximation to overcome the usual handicap of a severe spherical
harmonics truncation. A powerful iterative solver then allows us to
compute the frequencies of the most promising low-order modes using as
many spherical harmonics as necessary. The induced $\od{1\%}$ errors of
our model are now within observational limits. A plausible
seismological model of Jupiter might thus be at hand provided the use
of a more realistic description of the planet interior.
}

\maketitle

\section{Introduction}

Detecting global Jovian oscillations is still a major challenge since,
even with the most recent detection techniques, no acoustic or gravity
modes have been convincingly identified (Schmider et~al. \cite{Sch91};
Mosser et~al. \cite{Mos93,Mos00}; Cacciani et~al. \cite{Cac01}).
Furthermore, the collision of the Shoemaker-Levy9 fragments was
unfortunately not energetic enough as to allow any positive detection
of primary waves (Walter et~al. \cite{Wal96}; Mosser et~al.
\cite{Mos96}). 

Upcoming observational projects, such as JOVIS (Baglin \& Mosser
\cite{Bag99}), open a new window on the planet dynamics with the
possibility of detecting long-period mode oscillations. Because of
Jupiter's rapid rotation, inertial modes would be good candidates. These
have first been investigated by Lee \& Saio (\cite{Lee90}) and Lee
et~al \cite{Lee92} (hereafter LSVH) using different models for the
planet. Their approach was nonetheless limited by the use of no more than
three spherical harmonics in the angular description of the solutions
(see \S\ref{specif}). We have today the numerical tools to tackle the
same problem by including as many terms as necessary in each of the
spherical harmonic expansions (Rieutord \& Valdettaro \cite{Rie97};
Dintrans et~al. \cite{Din99}).

The plan of the paper is as follows: in \S\ref{model} we introduce our
model, its specificity and the equations we solve for. In \S\ref{ident},
we discuss mode identification methods in presence of rotation and show
how the inclusion of diffusion allows us to discriminate the inertial
eigenmodes in terms of least/high damped ones. Our results are then
presented in \S\ref{results} before concluding in \S\ref{conclusion}.

\section{Our model}
\label{model}

\subsection{Specificity of our study}
\label{specif}

The properties of inertial mode oscillations are commonly investigated
by first expanding each perturbation in a serie of spherical
harmonics. However, because of the Coriolis couplings between the
harmonics $(\ell,\ell \pm 1)$, one is faced with an infinite system of
coupled ordinary differential equations. Such a system is challenging
in its coding and requires prohibitive memory. LSVH solved these two
aspects by considering at most the first three terms in each of the
spherical harmonic expansions. They nevertheless recognized the necessity
of calculations with larger numbers of spherical harmonics before
detailed comparisons could be made between the theoretically calculated
frequencies and the yet to be measured oscillations of Jupiter. Indeed,
for the higher overtones, {\it truncation errors as high as 30\%} were
reached when two harmonics instead of three were used.

In our case, the memory requirement is immediately dealt with by
adopting the anelastic approximation (see also Dintrans \& Rieutord
\cite{Din00}). The idea is to filter out the acoustic waves in the
infinite system while keeping the density variations across the planet.
This is further justifiable since such high-frequency waves are
irrelevant to the dynamics of the inertial modes. Remains the coding
aspect of the problem which is taken care of by a preprocessing Perl
program capable of managing as many spherical harmonics as necessary.
Errors in our approach are thus mainly induced by the use of the
anelastic approximation and are estimated to be $\od{1\%}$ (see
Dintrans \& Rieutord \cite{Din01}).

\subsection{The oscillation equations under the anelastic approximation}

Assuming the time-dependence of the eigenmodes to be $\exp i \sigma t$
(where $\sigma$ is the angular frequency in units of rad/s), the
anelastic equations for the linear perturbations in the co-rotating
frame are given by

\begin{equation}
\left\{ \begin{array}{l}
i \sigma \vec{u} + 2 \vec{\Omega} \times \vec{u} = - \vec{\nabla} 
(P'/\rho_0), \\ \\
\Div (\rho_0 \vec{u} ) = 0,
\end{array} \right. 
\label{syst1}
\end{equation}

\noindent where $\vec{u}$ is the velocity, $P'$ the eulerian pressure
perturbation, $\Omega$ the angular frequency ($2 \vec{\Omega} \times
\vec{u}$ being the Coriolis force) and $\rho_0$ denotes the equilibrium
density of our interior model of Jupiter which we describe by the $n=1$
polytrope (Hubbard \cite{Hub84}). As boundary conditions, we impose the
regularity of the velocity at the centre while the anelastic approximation
implies that the radial component of the velocity vanishes at the surface
(see Dintrans \& Rieutord \cite{Din01}).

Using the spherical coordinate system $(r,\theta,\varphi)$, we expand
the velocity on spherical harmonics $Y^m_\ell (\theta,\varphi)$ (Rieutord
\cite{Rie91}). One arrives at an infinite set of coupled {\it radial}
equations to be solved as a generalized eigenvalue problem of the form

\begin{equation}
{\cal M}_A \vec{\psi}_{m^\pm} = \lambda {\cal M}_B \vec{\psi}_{m^\pm},
\label{eigenv}
\end{equation}

\noindent where $\lambda = i \sigma$ is the complex eigenvalue associated
with eigenvector $\vec{\psi}_{m^\pm}$. The $m^+$-modes corresponds to
even solutions ($\ell = |m|,|m|+2,|m|+4,\dots$) whereas $m^-$-modes are
odd ones ($\ell = |m|+1,|m|+3,|m|+5,\dots$). We note that even (odd)
modes define solutions that are symmetric (antisymmetric) with respect
to the equatorial plane. Finally, this eigenvalue problem is discretized
on the Gauss-Lobatto grid associated with Chebyshev polynomials yielding
to matrices ${\cal M}_A$ and ${\cal M}_B$ of order about $L\times (N+1)$
when $L$ spherical harmonics and $N$ polynomials are considered.

\section{Mode identification in presence of rotation}
\label{ident}

In this section we discuss the notion of mode identification when
adding rotation to the $p$- and $g$-modes as compared to the case of
inertial modes.

\subsection{$p$- and $g$-modes}

One recalls that two cases need to be considered depending on the
non-rotating frequency shift $\sigma_{k\ell}$ ($k$ being the radial order)
with respect to the Coriolis frequency $2\Omega$:

\noindent i) $\sigma_{k\ell} \gg 2 \Omega$ ($p$- and low-order
$g$-modes):  the weak Coriolis couplings can be described by a
perturbative theory (i.e. $\sigma_{k\ell m} \simeq \sigma_{k\ell} + m
C_{k\ell} \Omega$) making the $(k,\ell,m)$ identification possible.

\noindent ii) $\sigma_{k\ell} \infapp 2 \Omega$ (high-order $g$-modes):
the strong Coriolis couplings prevent any $\ell$-identification
of modes. A pulsation mode cannot henceforth be identified by a set
$(\ell,m)$ since the description of its angular dependence in terms of
a single spherical harmonic $Y^m_\ell (\theta,\varphi)$ is not possible
(Dintrans et~al. 1999).

\subsection{Inertial modes}

Only the $\sigma_{k\ell} \le 2 \Omega$ regime exists, i.e. the spectrum
is bounded by the Coriolis frequency. The Coriolis couplings are
dominant over the entire spectrum making the $\ell$-identification
meaningless. Although they recognized this fact, LSVH chose ``for
simplicity'' to attribute a given $\ell$-value to their inertial
eigenmodes; i.e. they arranged their low-order solutions in terms of
$\ell=1,2,3,4$ for $m=-1,-2,$ and $m=-3$ with eigenvalues sorted in
decreasing values. This arrangement is in fact unappropriate, as shown
for example with the eigenmodes of the incompressible rotating sphere
(Bryan 1889): for $m=0^+$, the first four low-order frequencies are
$\sigma/2\Omega = [0.654;0.830;0.469;0.677]$ with no possible
$\ell$-attribution (see also Greenspan \cite{Gre69}).

\subsection{Identification from the $(|\hbox{\ntau}|,\sigma)$-diagram}

Low-degree and low-order $p$- or $g$-modes are commonly identified in the
computed spectrum by plotting the $(\ell,\sigma)$-diagram. Unfortunately,
such a tool cannot be used for inertial modes since $\ell$ is not a
suitable quantum number anymore.

However, one would expect that the most promising observational
inertial modes are the ones with the weakest Coriolis couplings. Such
modes, hereafter referred to as ``low-order'' modes, only involve a
small family of spherical harmonics and are for this reason associated
to the smallest resolutions $L$. Therefore the challenge is to
extract them among the entire computed modes defining the spectrum,
thus the question: ``At a given $m$ and frequency shift $\sigma$,
what is the inertial mode which involves the least of spherical
harmonics, that is, the one whom the $\ell$-couplings are the weakest
?''. As we will prove in \S\ref{truncation} and in more details in
a forthcoming paper, one way to extract the modes in question consists in
plotting a $(|\hbox{\ntau}|,\sigma)$-diagram, \ntau~being the (negative)
damping rate of the mode when adding a slight diffusion.

\section{Results}
\label{results}

Hereafter dimensionless angular eigenfrequencies $\widetilde{\sigma}$
and damping rates $\widetilde{\hbox \ntau}$ are given in units of
$2\Omega$, that is $\sigma = 2\Omega \widetilde{\sigma}$ and \ntau
= $2\Omega \widetilde{\hbox \ntau}$.  For Jupiter, it means that
oscillation frequencies are $\nu ({\rm mHz})= \sigma / 2\pi \simeq
0.056 \widetilde{\sigma}$ whereas periods are $T ({\rm hr})= 2\pi/\sigma
\simeq 4.958/\widetilde{\sigma}$.

\subsection{Low-order modes for $m=-1^+,0^+,1^+$}
\label{observ}

\begin{figure}
\psfrag{x}{$|\widetilde{\hbox \ntau}|$}
\psfrag{y}{$\widetilde{\sigma}$}
\psfrag{toto}{$\ell$-couplings}
\centerline{\includegraphics[width=0.4\textwidth]{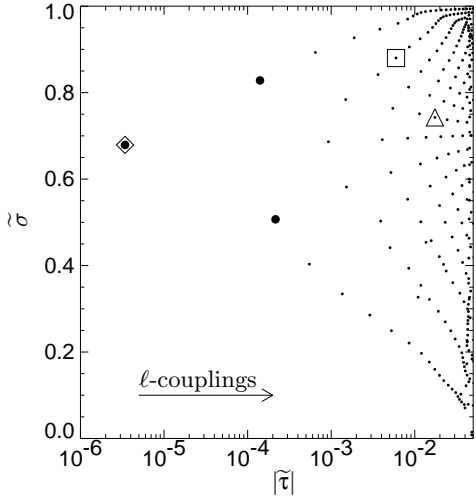}}
\caption{The $(|\widetilde{\hbox \ntau}|,\widetilde{\sigma})$-diagram
for the $m=0^+$ eigenvalues, where $\widetilde{\hbox \ntau}$ and
$\widetilde{\sigma}$ are the dimensionless damping rate and frequency,
respectively. The thick dots to the left denote the first three
low-order modes. The diamond, square and triangle modes, chosen in order of
increasing $\ell$-couplings (arrow), are used when discussing
truncation effects in \S\ref{truncation}.}
\label{distrib} 
\end{figure}

We add a small viscosity term to the momentum equation in
order to organize the eigenvalues in terms of least/high damped
pulsation modes. By plotting them in the $(|\widetilde{\hbox
\ntau}|,\widetilde{\sigma})$-diagram, we are able to identify the
low-order modes which position themselves in the left region of
the plot. An example is shown in Fig.~\ref{distrib} for the $m=0^+$
eigenvalues where few low-order modes are depicted by thick dots.
This figure was obtained using the QZ algorithm with the resolution
$L=N=60$ and an Ekman number $E=10^{-5}$ ($E=\nu/2\Omega R^2$, where $\nu$
is the kinematic viscosity and $R$ the Jupiter radius).

\begin{table}[b]
\caption[]{Dimensionless frequencies $\widetilde{\sigma}$ of the first five 
low-order modes for $m=-1^+,0^+,1^+$. The associated oscillation periods 
(in hours) are given in parenthesis.}
$$
\begin{array}{c@{\hspace{0.7cm}}c@{\hspace{0.7cm}}c}
\hline
\noalign{\smallskip}
-1^+ & 0^+ & 1^+ \\
\noalign{\smallskip}
\hline
\noalign{\smallskip}
0.656~(7.558) & 0.679~(7.302) & 0.706~(7.023) \\
0.201~(24.67) & 0.828~(5.988) & 0.863~(5.745) \\
0.799~(6.205) & 0.507~(9.779) & 0.719~(6.896) \\
0.491~(10.10) & 0.403~(12.30) & 0.921~(5.383) \\
0.327~(15.16) & 0.893~(5.552) & 0.343~(14.45) \\
\noalign{\smallskip}
\hline
\end{array}
$$
\label{table}
\end{table}

Once the best low-order modes isolated, remains the important task/step
of accurate computation of the corresponding eigenfrequencies in the
astrophysical relevant limit of zero viscosity. The QZ algorithm is not
the most suitable for this job because the matrix band structure of
$\cal M_A$ and $\cal M_B$ is not taken advantage of; i.e.
Fig.~\ref{distrib} required 1.8~Gbytes of memory and 16 hours of
computation on Origin2000.

We use an Arnoldi-Chebyshev iterative solver to extract the interesting
eigenvalues lost in the dense spectrum at $E=0$. Taking advantage of
the matrix band structure, it leads to substantial gain of both memory
and computing time. Table \ref{table} shows the computed frequencies of
the first five low-order modes for the $m=0^+$ and $m=\pm 1^+$ cases.
As can be noted from the associated periods given in parenthesis (in
hours), low-order inertial modes -- which must always exceed $T_{\rm
rot}/2 \sim 4.958$~hr -- can be as high as one day. It is thus clear
that these very long period modes require continuous observation using
either a ground-based network of dedicated telescopes or, yet better, a
devoted satellite such as the mentioned JOVIS project.

\begin{figure}
\psfrag{x}{$\ell$}
\psfrag{y}{\hspace{-1cm} Power spectrum}
\centerline{\includegraphics[width=0.4\textwidth]{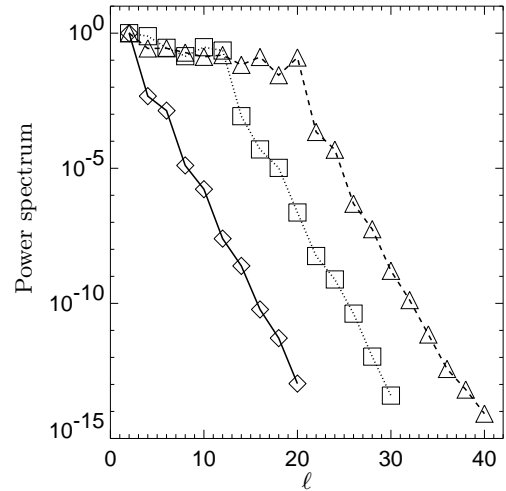}}
\caption{Normalized angular power spectra of the radial component $u_r$
of the velocity for the three modes labelled by a diamond, a square
and a triangle in Fig.~\ref{distrib}. Note the increase of the 
$\ell$-couplings with the mode order.}
\label{spec} 
\end{figure}

\subsection{Truncation effects}
\label{truncation}

In order to test the effect of the harmonic truncation, we compute, in the
limit $E=0$, the $\ell$-spectra of the three modes denoted by a diamond,
a square and a triangle in Fig.~\ref{distrib}. That is, for each $\ell$,
the maximum value over the Chebyshev coefficients of $u_r$ is plotted.

Fig.~\ref{spec} shows that only few spherical harmonics are necessary
to describe the lowest-order eigenmode (diamond) whereas many more
harmonics are required to resolve the square and triangle ones. To
reach a relative error of $\od{10^{-5}}$, we indeed checked that the
diamond mode only requires $L=4$ spherical harmonics while the square
and triangle modes necessitate at least $L=14$ and $L=22$ harmonics,
respectively. That is, eigenvalues are extremely sensitive to
$L$-values before the sharp drop in the curves of Fig.~\ref{spec}.

Combining our findings from Figs.~\ref{distrib} and \ref{spec}, the
link between the damping rate of an inertial mode and the strength of
its $\ell$-couplings is clearly demonstrated: the higher the damping
rates, the higher the $\ell$-couplings, thus the higher the needed
resolution $L$. It well justifies the use of the $(|\widetilde{\hbox
\ntau}|,\widetilde{\sigma})$-diagram to identify the low-order inertial
modes and to remedy to the fact that $\ell$ is no longer a suitable
quantum number.

\section{Conclusion} 
\label{conclusion}

We investigated the properties of the inertial modes of Jupiter for an
$n=1$ polytropic model of the planet. Our calculations have been performed
with as many spherical harmonics as necessary to resolve eigenmodes in
the anelastic approximation. 

We showed that adding a slight viscosity to the equations allows us to
identify what we referred to as ``low-order'' modes, that is inertial
modes which only involve a few spherical harmonics and are the most
promising for observations. Once these low-order modes identified, we
compute them in the relevant adiabatic limit of zero viscosity. We also
showed that, while the lowest-order mode is fully described by four
spherical harmonics, the following low-order modes require many more
harmonics (up to twenty) as a consequence of the succeeding Coriolis
couplings.

An avenue for future work consists on taking into account density
discontinuities in the planet's interior (such as the so-called PPT or the
plasma phase transition; Stevenson \& Salpeter \cite{Ste76}; Zharkov \&
Trubitsyn \cite{Zha76}) and the radiative window at the surface
(Guillot et~al. \cite{Gui94}).

\begin{acknowledgements} 
We thank Michel Rieutord and Beno\^{\i}t Mosser for helpful comments
and Lorenzo Valdettaro and Michel Rieutord for letting us use their
package Linear Solver Builder.
\end{acknowledgements}


\begin{thebibliography}{20}

\bibitem[1999]{Bag99} Baglin, A. \& Mosser, B. 1999, BAAS, 31 (DPS
meeting 31, 08.13)

\bibitem[1889]{Bry89} Bryan, G. 1889, Phil.~Trans.~R.~Soc.~London, 180, 187

\bibitem[2001]{Cac01} Cacciani, A., Dolci, M., Moretti, P. F., et~al.
2001, A\&A, 372, 317

\bibitem[1999]{Din99} Dintrans, B., Rieutord, M., \& Valdettaro, L. 1999,
J. Fluid Mech., 398, 271

\bibitem[2000]{Din00} Dintrans, B., \& Rieutord, M. 2000, A\&A, 354, 86

\bibitem[2001]{Din01} Dintrans, B., \& Rieutord, M. 2001, MNRAS, 324, 635

\bibitem[1994]{Gui94} Guillot, T., Chabrier, G., Morel, P., \& Gautier, D.
1994, Icarus, 112, 354

\bibitem[1969]{Gre69} Greenspan, H. P. 1969, The theory of rotating fluids
(Cambridge University Press)

\bibitem[1984]{Hub84} Hubbard, W. B. 1984, Planetary Interiors (Van
Nostrand Reinhold Co., New York)

\bibitem[1990]{Lee90} Lee, U., \& Saio, H. 1990, ApJ, 359, L29

\bibitem[1992]{Lee92} Lee, U., Strohmayer, T. E., \& van {Horn},
H. M. 1992, ApJ, 397, 674 (LSVH)

\bibitem[1993]{Mos93} Mosser, B., M\'ekarnia, D., Maillard, J. P.,
et~al. 1993, A\&A, 267, 604

\bibitem[1996]{Mos96} Mosser, B., Galdemard, P., Lagage, P., et~al. 1996,
Icarus, 121, 331

\bibitem[2000]{Mos00} Mosser, B., Maillard, J.~P., \& M\'ekarnia, D. 2000,
Icarus, 144, 104

\bibitem[1991]{Rie91} Rieutord, M. 1991, Geophys.~Astrophys.~Fluid~Dyn.,
59, 185

\bibitem[1997]{Rie97} Rieutord, M., \& Valdettaro, L. 1997, J. Fluid Mech.,
341, 77

\bibitem[1991]{Sch91} Schmider, F. -X., Fossat, E., \& Mosser, B. 1991,
A\&A, 248, 281

\bibitem[1976]{Ste76} Stevenson, D. J., \& Salpeter, E. E. 1976, In:
Gehrels T. (eds.) Jupiter (University of Arizona Press)

\bibitem[1996]{Wal96} Walter, C. M., Marley, M. S., Hunten, D. M.,
et~al. 1996, Icarus, 121, 341

\bibitem[1976]{Zha76} Zharkov, V. N., \& Trubitsyn, V. P. 1976, In:
Gehrels T. (eds.) Jupiter (University of Arizona Press)

\end{thebibliography}
\end{document}